\documentclass[12pt,preprint]{aastex}
\begin{document}
\title{On Dissipation inside Turbulent Convection Zones from 3D
        Simulations of Solar Convection}
\author{Kaloyan Penev, Dimitar Sasselov}
\affil{Harvard--Smithsonian Center for Astrophysics, 60 Garden St., Cambridge, MA 02138}

\author{Frank Robinson, Pierre Demarque}
\affil{Department of Astronomy, Yale University, Box 208101,
New Haven, CT 06520-8101}

	\begin{abstract}
	The development of 2D and 3D simulations of solar convection
	has lead to a picture of convection quite unlike the usually
	assumed Kolmogorov spectrum turbulent flow. We investigate
	the impact of this changed structure on the dissipation
	properties of the convection zone, parametrized by an
	effective viscosity coefficient. We use an expansion treatment
	developed by Goodman \& Oh 1997, applied to a numerical model
	of solar convection (Robinson et al. 2003) to calculate an effective
	viscosity as a function of frequency and compare this to
	currently existing prescriptions based on the assumption of
	Kolmogorov turbulence (Zahn 1966, Goldreich \& Keeley 1977).
	The results match quite closely a linear scaling with period, 
	even though this same formalism applied to a
	Kolmogorov spectrum of eddies gives a scaling with power-law index of 
	$5\over3$. 
	\end{abstract}

	Turbulent (eddy) viscosity
        is often considered to be the main mechanism responsible for dissipation 
	of tides and oscillations  in convection zones of cool stars and
	planets (Goodman \& Oh 1997, and references therein).
	Currently existing descriptions have been used, with varying
	success, to explain circularization cut-off periods for main
	sequence binary stars (Zahn \& Bouchet 1989, Meibom \& Mathieu
	2005), the red edge of the Cepheid instability strip (Gonczi
	1982) and damping of solar oscillations (Goldreich \& Keeley 1977).
        However, this hypothesis has been far more successful in damping oscillations
	than damping tides, and different mechanisms have been proposed for the
	latter, especially for planets (see Wu 2004ab; Ogilvie \& Lin 2004, and references 
	therein). In this paper we reconsider the problem of tidal dissipation in stellar convection
	zones of solar-type stars using the turbulent velocity field from a
	realistic 3D solar simulation. \\
	
        The standard treatment is to  assume
	a Kolmogorov spectrum in the convection
	zone and apply some prescription to model the effectiveness of  eddies
	in dissipating the given perturbation. Two prescriptions have
	been proposed to describe the efficiency of eddies in  dissipating
	perturbations with periods smaller than the eddy turnover
	time.\\

	Firstly according to Zahn(1966, 1989), when the period of the
	perturbation(T) is shorter than the eddy turnover time 
	($\tau$) the dissipation efficiency is decreased because in half
	a period the eddy only completes  $T \over 2\tau$  of its
	churn, and hence the dissipation (viscosity) should be 
	inhibited by the same factor:
	\begin{equation}
		\nu = \nu_{max} \min\left[ \left(T\over2\tau\right),
		1\right]
	\end{equation}
	Where $\nu_{max}$ is some constant which depends on  the mixing length
	parameter. With this assumption large eddies dominate the
	dissipation. This prescription has been tested against  tidal
	circularization times for binaries containing a giant star
	(Verbunt and Phinney 1996), and is in general agreement with
	observations.\\

	Secondly, Goldreich \& Nicholson (1989) and Goldreich \& Keely
	(1977) argue that the viscosity should be severely suppressed
	for eddies with $\tau\gg T$, and hence the dissipation should
	be dominated by the largest eddies with turnover times less
	than $T/2\pi$. From Kolmogorov scaling the viscosity on a
	given time-scale is quadratic in the time-scale, or:
	\begin{equation}
		\nu = \nu_{max} \min\left[\left( T\over 2\pi\tau\right)^2,
		1\right]
	\end{equation}
	This description has been used successfully by Goldreich \& 
	Keely (1977), Goldreich \& Kumar(1988), Goldreich, Kumar \& 
	Murray (1994) to develop a theory for the damping of the 
	solar $p$-modes. If the more effective dissipation was applied
	instead, severe changes would be required in the excitation
	mechanism in order to explain the observed  $p$ mode
	amplitudes. However, this inefficient dissipation is
	inconsistent with observed tidal circularization for binary
	stars (Meibom \& Mathieu 2005). Additionally, Gonczi (1982)
	argues
	that for pulsating stars the location of the red edge of the
	instability strip is more consistent with Zahn's description
	of eddy viscosity than with that of Goldreich and
	collaborators.\\

	However, Goodman \& Oh (1997) gave a consistent hydrostatic
	derivation of the convective viscosity, using a perturbational
	approach. For a Kolmogorov scaling they obtained a result that is
	closer to the less efficient Goldreich \& Nicholson viscosity than it is to 
	Zahn's. While providing a more sound theoretical
	basis for the former scaling, this does not resolve the
	observational problem of insufficient tidal dissipation.\\ 

	Both 2D and 3D numerical simulations of the solar convection
	zone have revealed that the picture of a Kolmogorov spectrum
	of eddies is too simplified (Stein \& Nordlund 1989, Robinson
	et al. 2003). The  simulations showed
	that convection proceeds in a rather different,  highly
	asymmetric fashion. This suggests that the problem of
	insufficient dissipation may be resolved by replacing the
	assumption of Kolmogorov turbulence with the velocity
	field produced from numerical simulations. More importantly,
	an asymmetric and non-Kolmogorov turbulence might dissipate
	different perturbations differently, i.e. depending both on
	the frequency and geometry of the perturbation. Such 
	simulations have been used to develop a better model for the 
	excitation of solar $p$-modes (Samadi et al. 2003).\\

	Our approach is to apply the Goodman \& Oh (1997)
	formalism to the velocity field obtained from  realistic  3D solar surface convection 
	in a  small box. The 3D simulation was  
	able to  reproduce the frequency spectrum of
	solar $p$-modes. The main result is that we find a scaling relation with
	frequency that is in better agreement with the more
efficient scaling proposed by Zahn, albeit for different reasons.

\section{Method}
	We apply the Goodman \& Oh (1997) treatment of convection 
	to the velocity field of a 3D simulation of the outer layers of the sun.
        Goodman \& Oh  
	assume that a steady state convection zone velocity
	field ($\mathbf{v}$) is
	perturbed by introducing an external velocity ($\mathbf{V}$). They also assume
	that the convection occurs on scales small compared to the
	perturbation, and further that the convection is approximately
	incompressible and isentropic. Assuming that the convective
	length scales are small compared to the perturbation allowed
	them to consider a volume small enough to 
	accommodate all convective scales, but over that volume the
	perturbation velocity field can be assumed linear in the
	Cartesian coordinates ($\mathbf{x}$):
	\begin{equation}
		\mathbf{V} = \mathbf{A}(t)\cdot \mathbf{x}
	\end{equation}
        In other words we define the matrix $\mathbf{A}$ as the
        derivative matrix of $\mathbf{V}$:
        \begin{displaymath}
                A_{i,j} = \frac{\partial \mathbf{V}_i}{\partial x^j}
        \end{displaymath}
%
%  Should j be a subscript on x ?
        And keep only the first term in the Taylor series of
        $\mathbf{V}$.\\

	Under this assumption means  the results will only be applicable
	to perturbations that are large compared to the size of the simulation domain.
	In particular this prevents us from making any statements
	about the 5 minute solar oscillations, because the
	penetration depth of those is less than the box we use, and
	the coarse resolution prevent us from looking at only the
	upper part of the box.\\
	
	Assuming incompressible and isentropic convection allows one
	to use the Eulerian equations for fluid motion:
	\begin{eqnarray}
		&\partial_t\mathbf{v} +
		\mathbf{V}\cdot\nabla\mathbf{v} +
		\mathbf{v}\cdot\nabla\mathbf{V} +
		\mathbf{v}\cdot\nabla\mathbf{v} + \nabla w=0&
		\label{eq: Euler 1}\\
		&\nabla\cdot\mathbf{v}=0 \label{eq: Euler 2},&
	\end{eqnarray}
	where $\nabla w$ incorporates pressure and gravitational
	acceleration, assumed to be gradients of scalar fields.\\

	The problem has two dimensionless parameters: the tidal strain
	$\Omega^{-1} \left|\mathbf{A}\right|$, and $\left(\Omega
	\tau_c\right)^{-1}$, where $\Omega$ is the frequency of the
	perturbation and $\tau_c\equiv \frac{L_c}{V_c}$. The characteristic convective length scale
       is  $L_c$ and $V_c$  is the   
	characteristic convective velocity. In the case of
	hierarchical eddie structured convection $\tau_c$ is the eddy
	turnover time.\\

	So using eq. \ref{eq: Euler 1} and eq. \ref{eq: Euler 2} one 
	can express the perturbation in the convection velocity field in a 
	coordinate system moving with the perturbation. Expanding in
	powers of the above dimensionless parameters and keeping only
	first order terms gives:
	\begin{equation}
		\delta_{1,1} \mathbf{v'}(\mathbf{k}, \omega)
		=-\frac{i}{\omega}\mathbf{P_k}\cdot
		\left[\mathbf{A}(\Omega)\cdot\mathbf{v}_0(\omega-\Omega,\mathbf{k})+
		\mathbf{A}(-\Omega)\cdot\mathbf{v}_0(\omega+\Omega),\mathbf{k})\right]
		\label{eq: delta v}
	\end{equation}
	The subscripts of $\delta_{1,1} \mathbf{v'}(\mathbf{k},
	\omega)$ indicate that only first order terms in the
	dimensionless parameters have been included, primes
	indicate quantities expressed in a coordinate system moving
	with the perturbation, and $\mathbf{v}_0$ is the convective
	velocity field in the absence of the perturbation. All of the
	above quantities are in Fourier space, because there the
	incompressibility is simply imposed by the projection
	operator:
	\begin{displaymath}
		\mathbf{P_k}\equiv\mathbf{I}-\frac{\mathbf{kk}}{k^2}
	\end{displaymath}
	Eq. \ref{eq: delta v} can then be used to express the energy
	dissipation rate again as a power series in the two
	dimensionless quantities.  Goodman and Oh s' treatment 
	implicitly assumes the box is small enough
	for the density not to vary significantly, and so 
	it is sufficient to write the energy per unit mass as $\left<
	\mathbf{v}\cdot\mathbf{v}\right>$ and assume that to be   
	independent of position.\\

	In our case the simulation encompasses about 8  pressure scale heights 
        so that the density varies significantly between 
	the top and bottom. This means   
	we need to use the
	dissipation per unit volume -
	$\left<\rho\mathbf{v}\cdot\mathbf{v}\right>$ - instead.\\

	In order to avoid taking a 7 dimensional integral, which would
	be prohibitive in terms of computation time, we replace  
	the density with its horizontal and temporal average
	leaving  only the most important vertical dimension.
	Taking the time derivative of the energy per unit volume using
	that density and the perturbed convective velocity, our
	expression for the rate of dissipation per unit volume
	to lowest order becomes:
	\begin{eqnarray}
		\dot{\mathcal{E}}_{2,2}=
		\mathbf{Re}\Bigg\{ \int\frac{d^3\mathbf{k}\;dk'_z}{(2\pi)^4}
		\rho^*(k_z+k'_z)\Big[ \left<
		\mathbf{v}_0(\mathbf{k},-\Omega)\cdot\mathbf{A}(\Omega)\cdot
		\mathbf{P_{k'}}\cdot\mathbf{A}(\Omega)\mathbf{v}_0(\mathbf{k'},
		-\Omega)\right> &&\nonumber\\
		+ \left.\mathbf{v}_0(\mathbf{k},-\Omega)\cdot\mathbf{A}(\Omega)\cdot
		\mathbf{P_{k'}}\cdot\mathbf{A}(-\Omega)\mathbf{v}_0(\mathbf{k'},
		\Omega)\right> \Big] \Bigg\}&&
		\label{eq: E dot full}
	\end{eqnarray}
	Where $k'=(-k_x, -k_y, k'_z)$, and the subscripts, as before, denote 
	the order in the two dimensionless parameters characterizing the tide 
	and the convection respectively. $\rho(k_z)$ is the fourier transform
	of the density averaged over $x,y,t$. The normalization is such that 
	$\rho(0)$ is the average density over all space and time.\\

	Eq. \ref{eq: E dot full} gives an anisotropic viscosity, for which
	we can obtain
	the different components by setting all terms of $\mathbf{A}$
	to $0$ except for one, and comparing to the equivalent
	expression for the molecular viscosity:
	\begin{equation}
		\dot{\mathcal{E}}_{visc} = \frac{1}{2}
		\left<\rho\nu\right>
		\;Trace\left[\mathbf{A}(\Omega)\cdot\mathbf{A^*}(\Omega)\right]
		\label{eq: E dot mol}
	\end{equation}
	Where the average is over the volume and over time. 
%Frank Robinson
\section{Realistic 3D solar surface convection}

The 3D simulation of the Sun is case D in Robinson et al. (2003). This
has dimensions 2700 km  $\times$ 2700 km  $\times$ 2800 km
on a $58 \times 58 \times 170$   grid.
A  detailed one-dimensional (1D)
evolutionary model e.g. see Guenther et al. (1992)
provided  the starting model for the
3D simulation. Full details of the
numerical approach and physical assumptions
are described in Robinson et al. (2003).

The simulation extended from a few hundred km above the photosphere down to a
depth of about 2500 km below the
visible surface (photosphere). This is about 8 pressure scale heights.
The box had  periodic side walls and impenetrable top and bottom surfaces with a 
constant energy flux fed into the base and a conducting top boundary.
The flux was computed from the 1D stellar
model, thus  was not arbitrary, but
was the correct amount of energy flux the computation domain should transport
outward in a particular star.

To get a thermally relaxed system in a reasonable amount of computer time,
they  used an implicit numerical scheme, ADISM (Alternating Direction Implicit
on a Staggered Mesh) developed by Chan \& Wolff (1982).
Careful attention was paid to the geometric size of the box. Importantly the domain was
deep enough and wide enough
to ensure the boundaries had  minimal effect on the bulk of the overturning
convective eddies
(or on the flow statistics).
The convection simulation was run using the ADISM code until it 
reached a statistically steady state. This was checked by 
confirming that the influx and outflux of  the box were  within 5 \% of
each other and the run of the maximum velocity
have reached an  asymptotic state.

\begin{figure}[tbp]
\begin{center}
	\includegraphics[angle=270,width=0.4\textwidth]{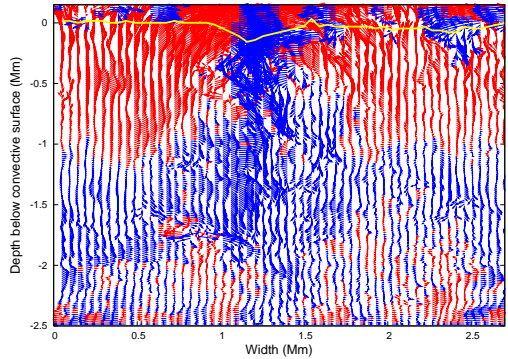}
	\caption{A sample snapshot of the convective flow.
	Blue color indicates downwarad flow, red indicates upward
	flow. The arrows show the
	velocity normalized to the sound speed. The yellow
	line represents the convective surface (i.e. where the
	entropy gradient is 0).
	}
	\label{fig: snapshot}
\end{center}
\end{figure}

After the model was relaxed they sampled the entire 3D velocity field at 1
minute intervals.
The data set used in this paper consists of  150 minutes of
such  solar surface convection. This is  about 20 granule turnover
life times. An example velocity snapshot of the convective flow is
presented in fig. \ref{fig: snapshot}.

% Frank robinson

\section{Results}
	We implement eqs. \ref{eq: E dot full} and \ref{eq: E dot mol}
	by taking discrete Fourier transforms (FFT) of the velocity
	field and the averaging density horizontally and over time.
	In doing so it is important to verify that the
	windows introduced by the limited time and space extent of the
	simulation box do not dominate the results. This was done by
	repeating the calculation with the raw results, without any
	windowing and with Welch and Bartlett windows applied to all
	the dimensions simultaneously. As expected this has little or 
	no effect on the frequency scaling (see below).\\ 

	\begin{figure}[tbp]
	\begin{center}
%		\plottwo{f2a}{f2b}
%		\plotone{f2c}
		\includegraphics[width=0.45\textwidth]{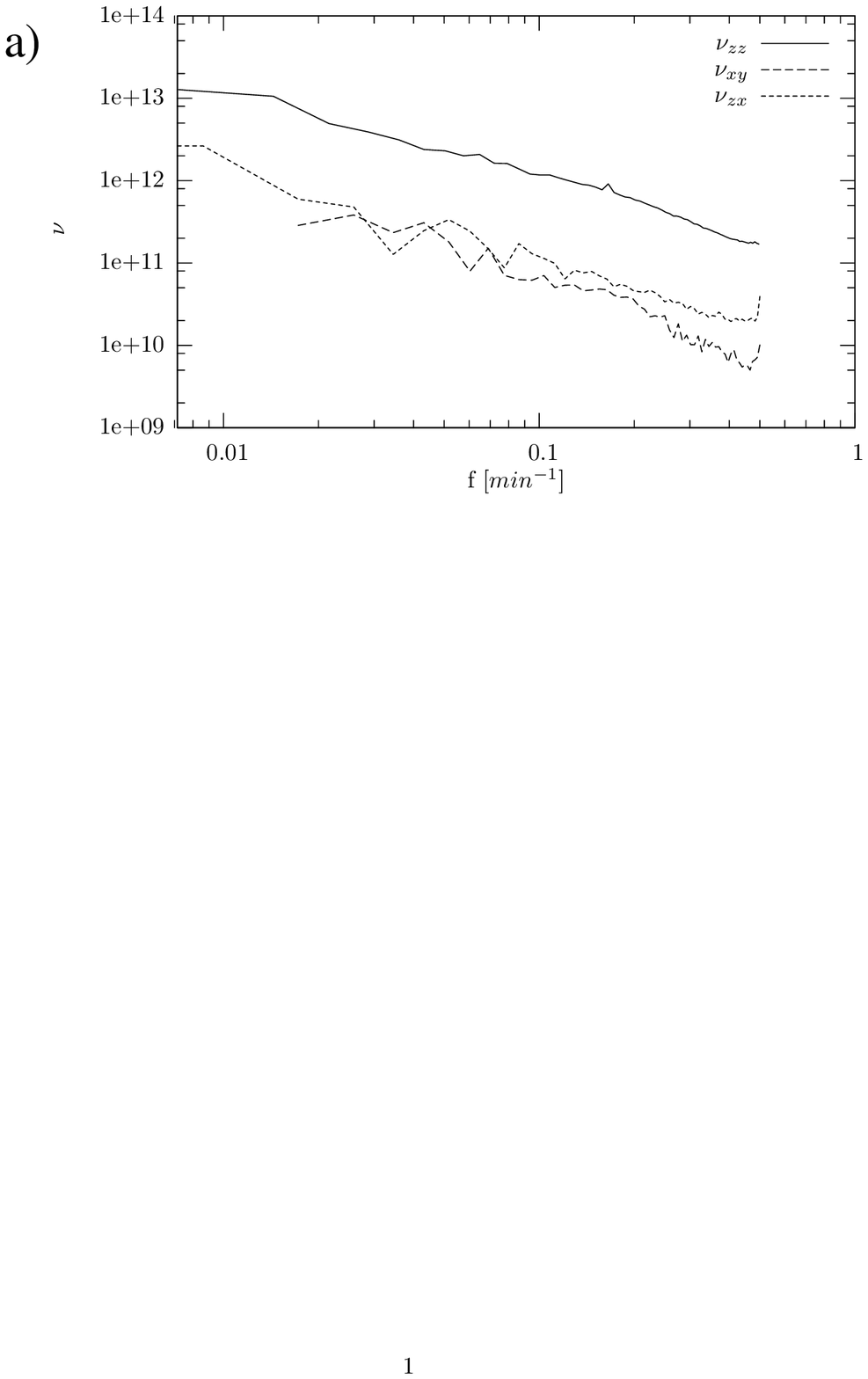}
		\includegraphics[width=0.45\textwidth]{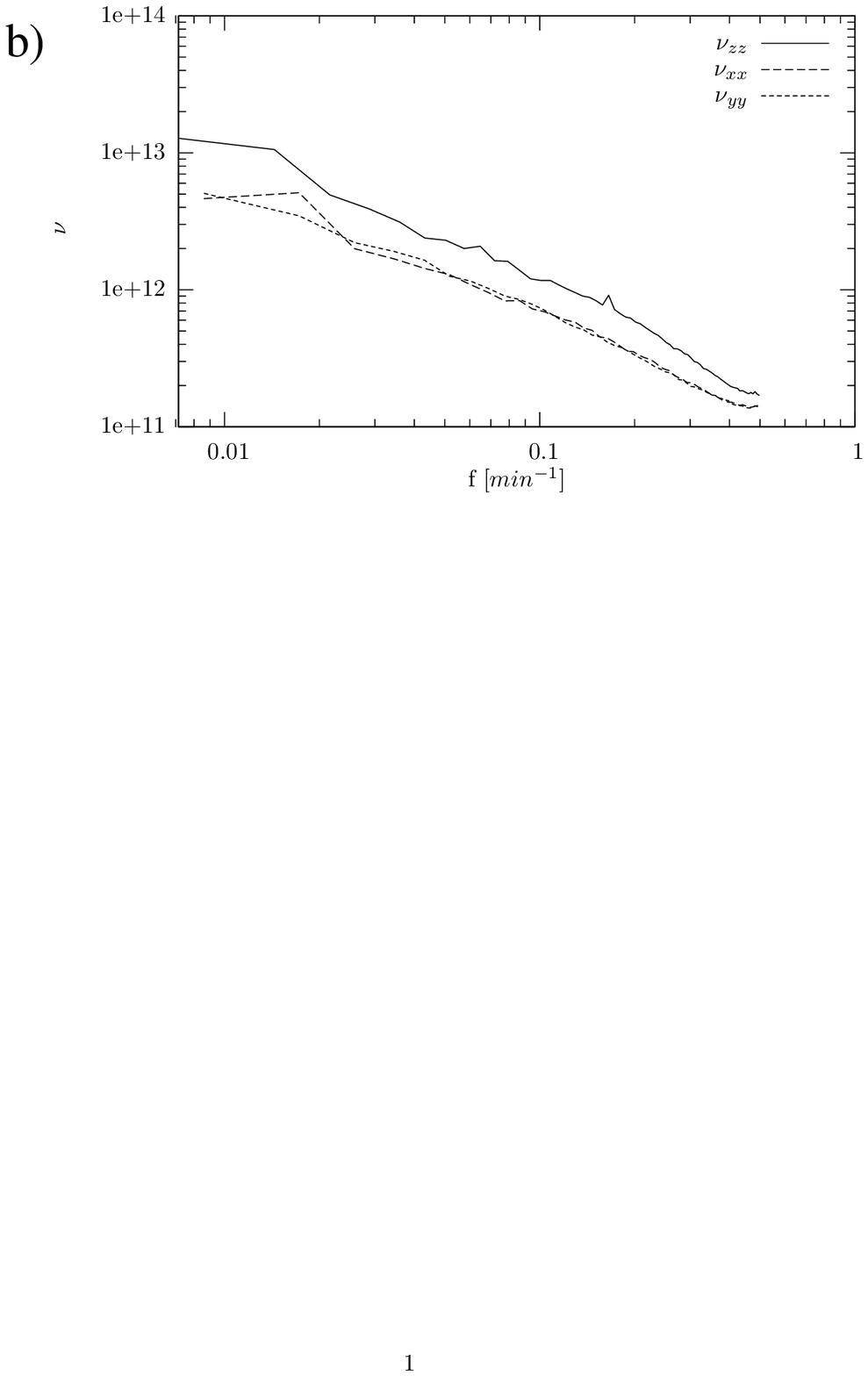}
		\includegraphics[width=0.45\textwidth]{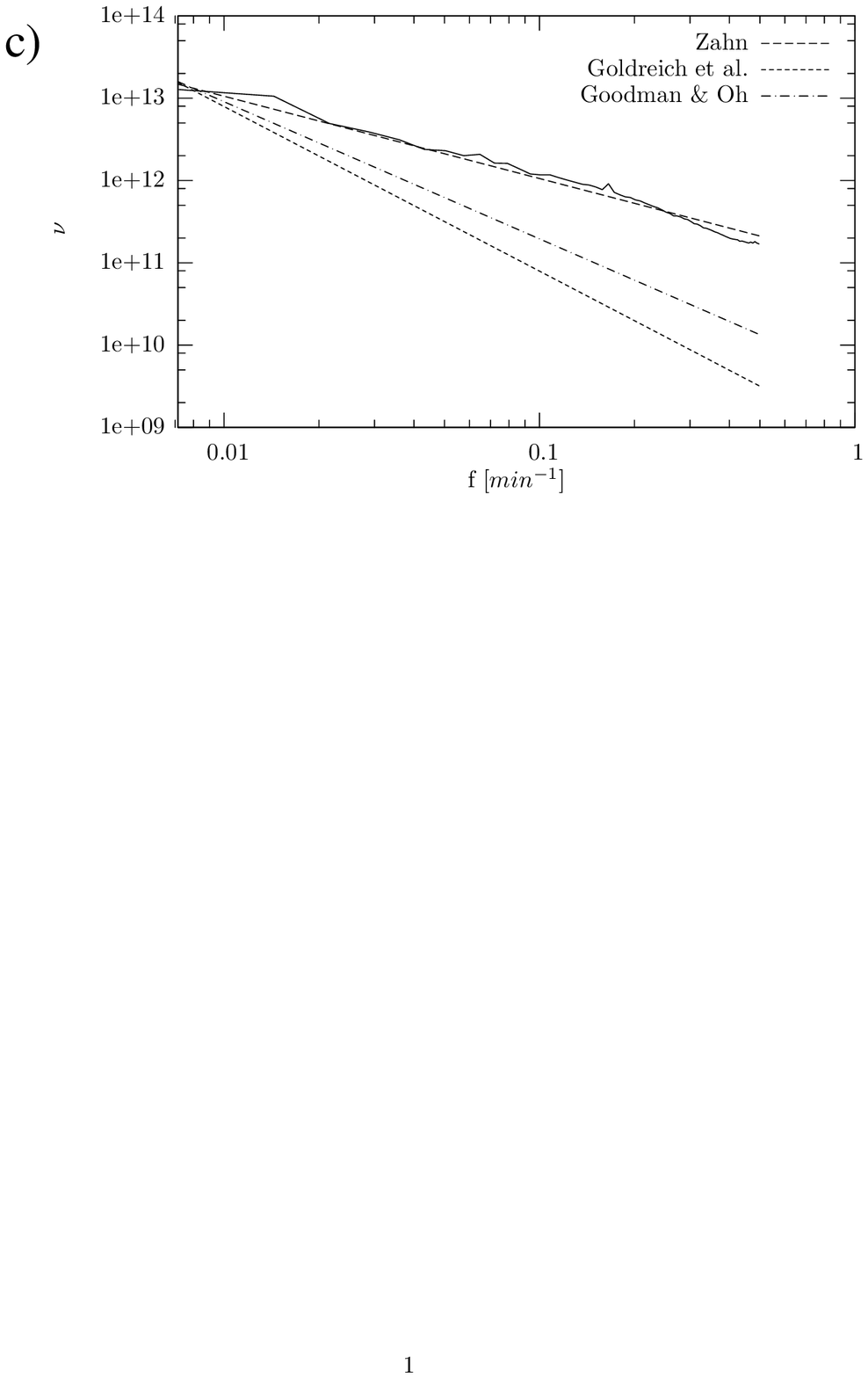}
		\caption{
		a) The off diagonal terms of the viscosity
		tensor compared to the z-z component.
		b) The diagonal terms of the viscosity tensor.
		c) The z-z component of the viscosity tensor (solid line)
		computed using eq. \ref{eq: E dot full} and eq. 
		\ref{eq: E dot mol} compared to the frequency 
		scalings proposed by Zahn, Goldreich et. al.
		and Goodman and Oh. 
		The horisontal axis for all the plots is 
		the frequency in cycles per min.
		}
		\label{fig: scalings}
	\end{center}
	\end{figure}

	As the viscosity tensor defined by eqs. \ref{eq: E dot full} and
	\ref{eq: E dot mol} is clearly symmetric, it only
	contains 6 independent real valued components. Figure
	\ref{fig: scalings} displays the values of the viscosities we calculated.

% you may need to state the expressions as it is not clear 

	Fig. \ref{fig: scalings}a shows that the off
	diagonal terms are completely insignificant compared to the
	diagonal terms. Since in all the situations that concern
	us, the divergence of the perturbation field is never small
	compared to the other derivatives of the perturbing velocity
	field, the dissipation will be dominated by the diagonal terms.
	Hence their scaling with frequency will determine how the
	dissipation scales. \\

	Fig. \ref{fig: scalings}b shows that all the
	diagonal components scale roughly the same way with frequency
	and are dominated by the z-z component, although
	not by that dramatic a difference. Furthermore, for perturbations
	like tides the z derivative of the z component of the
	perturbation velocity is the largest element of the matrix
	$\mathbf{A}$ and hence that will be the term that will
	determine the frequency scaling of the dissipation.\\
% Figures need to be in order 

	In Fig. \ref{fig: scalings}c we see the comparison between the
	different scalings with frequency suggested so far. We also show the
	scaling that we obtain by applying the Goodman and Oh (1997) method
	to a simulated 3D convection velocity field. The lines shown are 
	least square fits to the curve we obtain from the simulation 
	velocities. They seem to all intersect at the upper right-hand corner
	because the fits were done in linear space, not logarithmic, and 
	hence do not tolerate even small deviations in the upper portion 
	of the log-log plot. The best fit slope for our curve (not
	shown) is: 
	\begin{displaymath}
		\nu\propto \Omega^{1.1\pm0.1}
	\end{displaymath}
	regardless whether we do the fit in linear or 
	logarithmic space.\\

\begin{figure}[tbp]
	\begin{center}
		\includegraphics[width=0.45\textwidth]{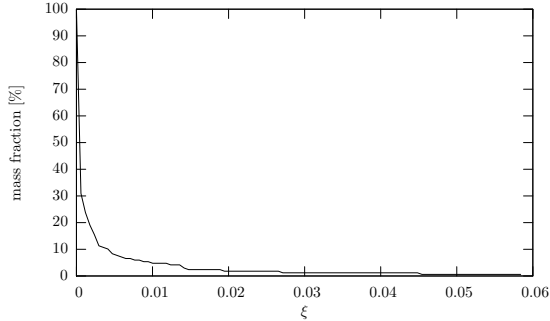}
		\caption{The fraction of the total mass residing in a
		region with compressibility parameter
		$\xi\equiv\tau_c\nabla\cdot\mathbf{v}$ less than the
		given value. 
		}
		\label{fig: M(xi)}
	\end{center}
	\end{figure}

	What are the  possible sources of
	error in this result? Firstly we have assumed an
	incompressible flow in order to simplify the treatment.
	However, the fluid simulations used are not incompressible,
	because at the top of the convection zone, where most of the
	driving of the convection occurs, the flow velocities reach
	very close to the speed of sound and hence the flow is
	necessarily compressible. However, even though that layer is
	extremely important for the flow established below, it only
	contributes insignifficantly to the turbullent dissipation,
	because it only contains a few percent of the total mass. 
	To
	verify that only a small fraction mass lies in a compressible
	region for each grid point, we define a compressibility
	parameter $\xi \equiv
	\tau_c\left|\nabla\cdot\mathbf{v}\right|$, where $\tau_c$ is
	the eddy turnover time in our box. In fig. \ref{fig: M(xi)} we
	plot the mass fraction with $\xi$ less than certain value. It
	is clear that the incompressibility assumption is violated
	only for a negligible fraction of the mass. As we noted before
	the flow is compressible only near the top of the box. To
	confirm that the presence of this region does not
	signifficantly affect our results we repeated the analysis
	separately for the top and bottom halves of the simulation
	box. The two new scalings obtained this way were completely
	consistent with the scaling of viscosity with frequency for
	the entire box.\\

	Next, the fact that we have a finite
	(small) portion of the convection zone, both in time and in
	space could be important. We only treat the top portion of the
	solar convection zone and hope that the result is not very
	sensitive to depth. Of course it would be ideal to have the
	entire depth of the convection zone covered, but with current
	computational resources this is way outside of reach.\\
	
	The finite span of the simulations may also be  introducing edge
	effects which can be treated by applying some sort of a window
	function. We tried Welch, Bartlett and square
	window (no window). To verify that the time window 
	available is large enough, we tried ignoring the last approximately
	$1/3$ of the data. We carried all those test on two independent
	runs of the model. The slopes this produced ranged from 
	$\nu\propto \Omega^{0.98}$ to $\nu\propto \Omega^{1.19}$, where most of
	the difference originated from the two independent runs.\\
	
	In addition the finite resolution
	might be leading to aliasing that could change our result.
	In particular make it flatter than it really is, by basically
	dumping additional power to the frequencies for which the
	dissipation is smallest (the places with higher value of the
	dissipation are less likely to be affected significantly).
	The effects of this can be seen in the diagonal viscosity
	components. The tails of their curves become flatter toward
	the end. The fact that this is restricted to the end of
	the curves is encouraging as  it suggests only the high frequency
	end of the curve is affected. Also we have looked at
	crossections of the Fourier transformed velocity field and
	they do tail off at high $|k|$, which gives us confidence
	that the resolution is sufficient to capture most of the
	spectral power and that aliasing effects will be small.\\

	Finally there are statistical errors associated with every
	point. Those can be estimated by noting the difference between
	$\nu_{xx}$ and $\nu_{yy}$ in Fig. \ref{fig: scalings}b.
	Physically one expects that there should be no differences
	between the two horizontal directions of the simulation box,
	so the differences between them is some sort of measure of
	the error. In particular from there one can see that the first
	few points (at the low frequency end) are significantly less
	reliable than the rest, but apart from the first few points
	those errors become small. The average fractional uncertainty is
	$\sim 3\%$, which leads to an overall error in the slope of
	$0.01$.\\ 

\begin{figure}[tb]
	\begin{center}
%		\plottwo{f4a}{f4b}
%		\plotone{f4c}
		\includegraphics[width=0.45\textwidth]{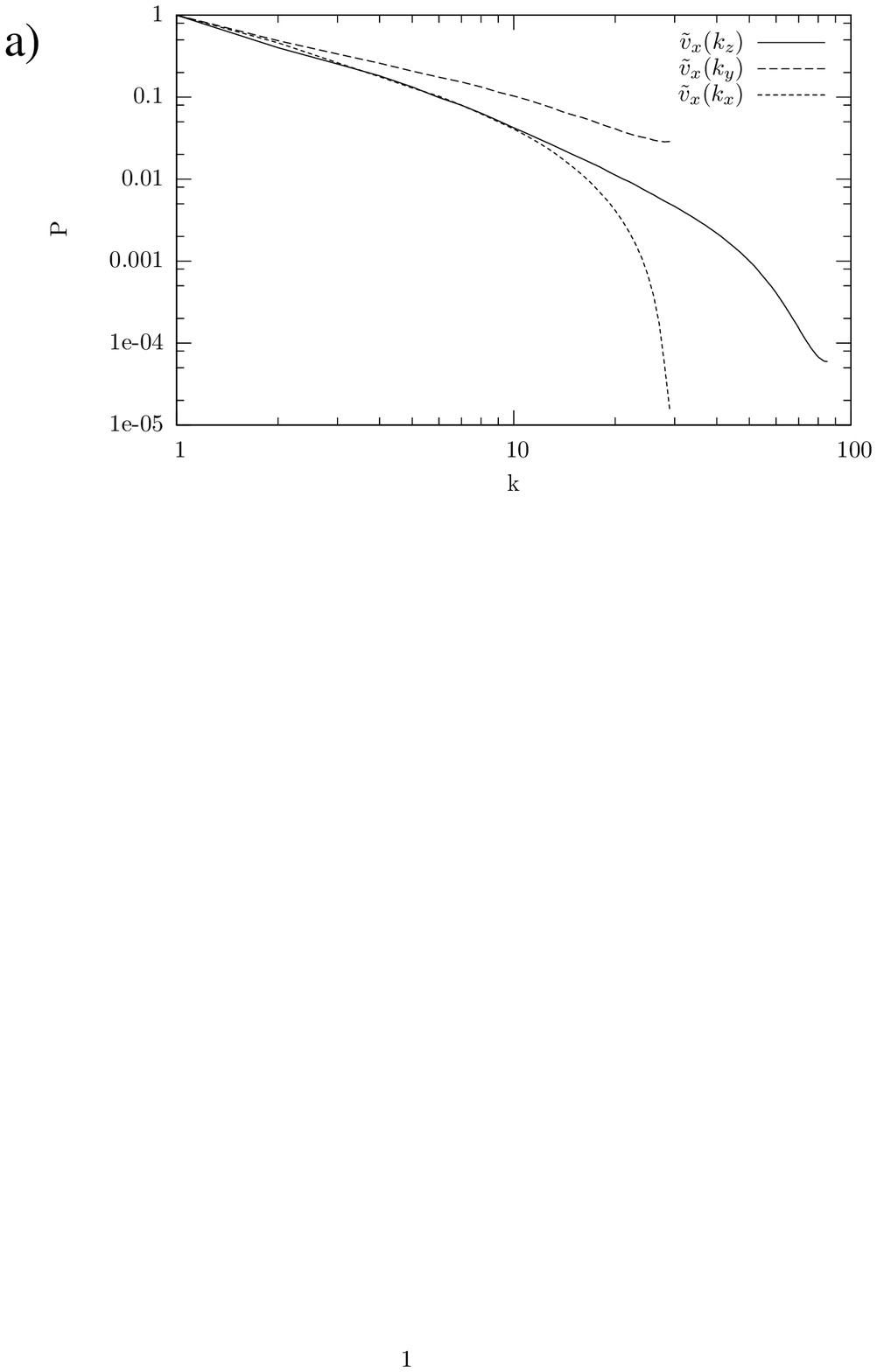}
		\includegraphics[width=0.45\textwidth]{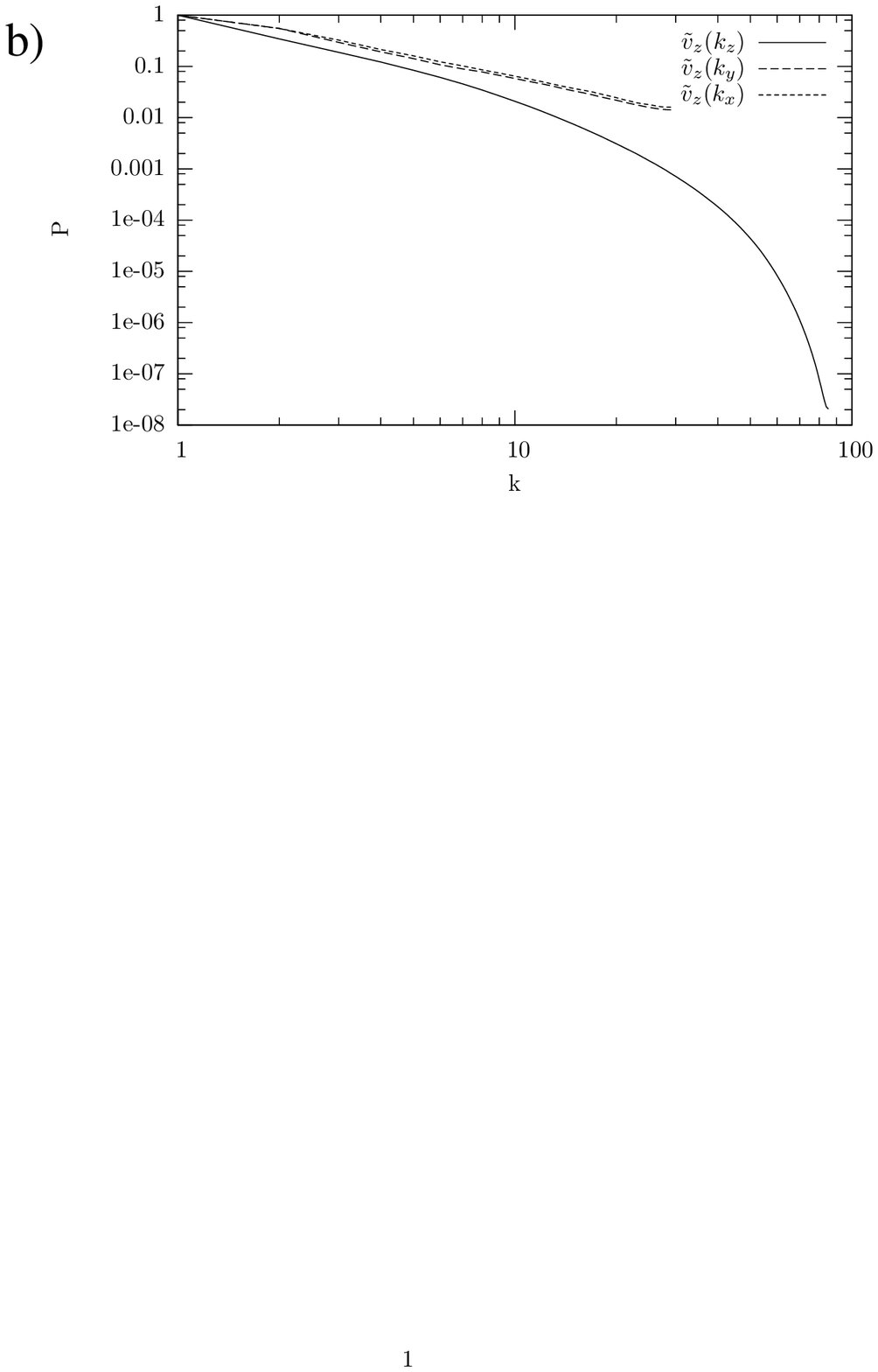}
		\includegraphics[width=0.45\textwidth]{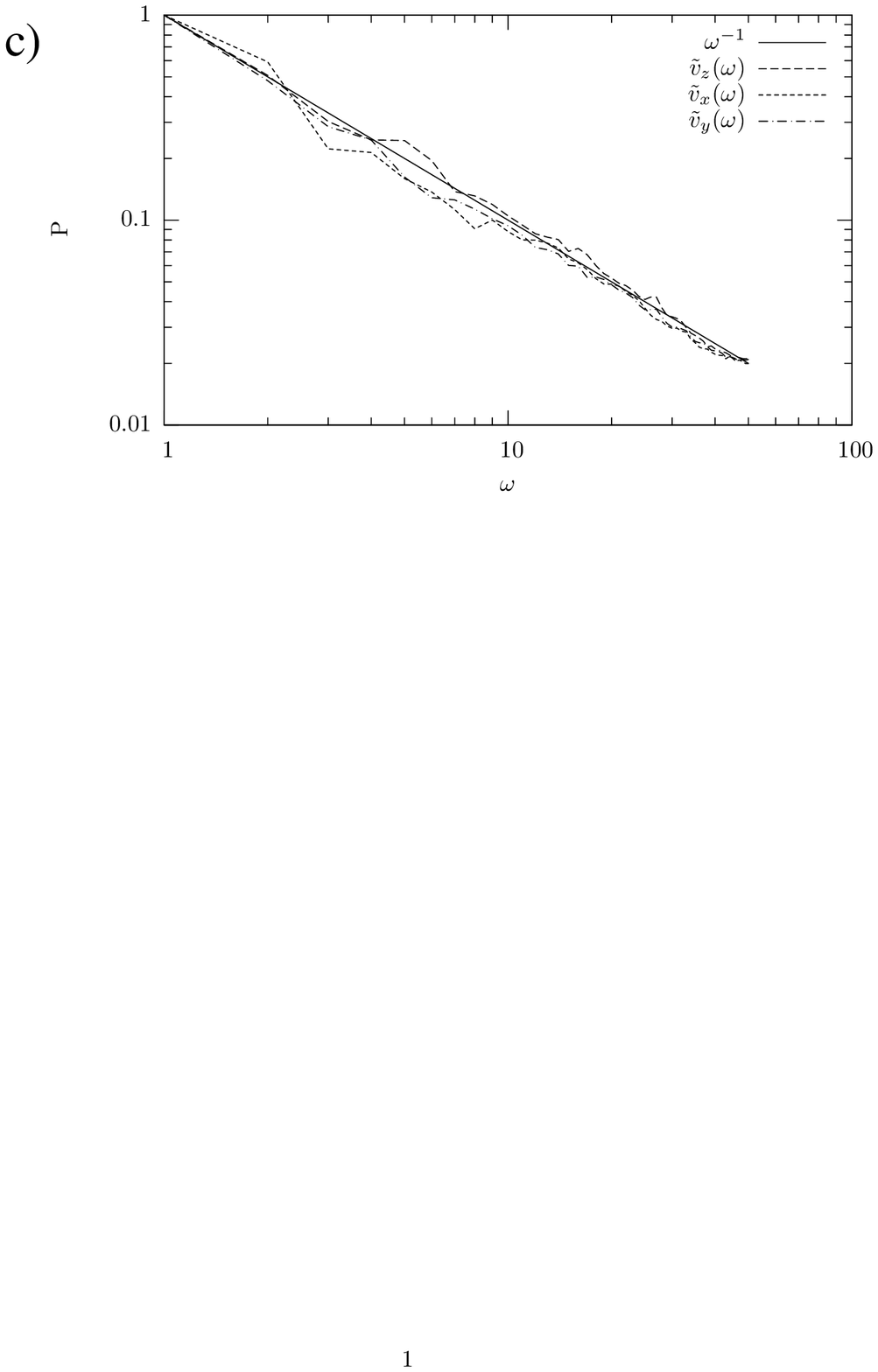}
		\caption{a) Spatial power spectrum of the
		horizontal velocities. Only
		one of the horizontal componenents is plotted but the
		power spectrum of the other horizontal component is 
		identical. 
		b) Spatial power spectrum of the radial velocity. 		
		c) Frequency power spectrum of the three
		velocity components. The straight solid line -
		$P\propto\omega^{-1}$ - gives a good approximation to
		all three scalings.}
		\label{fig: power spectra}
	\end{center}
	\end{figure}

	Abandoning the
	Kolmogorov picture of turbullence clearly has a large effect on the result.
	Even though we use the approach of Goodman \& Oh,
	which gives a power law index of $5/3$ for a Kolmogorov
	turbulence, our results give a scaling, rather different 
	from the previous prescriptions. We also find that the viscosity 
	is no longer isotropic. This is due to the signifficant
	difference in scaling between the velocity power spectrum with
	frequency and wavenumber in our simulation and the Kolmogorov
	prescription (see fig. \ref{fig: power spectra}). 
	There are two important distictions apparent.
	First the frequency spectrum of our box is much shallower than
	the Kolmogorov prescription. This is
	responsible for the slower loss of efficiency of viscosity
	with frequency that we observe. Second the radial direction is
	clearly very different from the two horizontal directions ---
	$v_x$ and $v_y$behave very differently from $v_z$ and the
	dependence of $\mathbf{v}$ on $x$ and $y$ is different from
	the $z$ dependence (fig. \ref{fig: power spectra} a, b) ---
	of course this results in the anisotropy of the viscosity
	tensor we calculate. Even though the spatial dependence of the
	horizontal velocity components is much different from the
	radial velocity spatial dependence, the frequency power
	spectrum of all three components scales roughly like
	$P\propto\Omega^{-1}$ (fig. \ref{fig: power spectra}c). From
	eq. \ref{eq: E dot full} we see that if all the components of
	$\mathbf{v}$ have the same scaling with frequency, that same
	scaling will also apply for the viscosity, which is indeed
	what we observe.\\

\section{Discussion}
	
	Our result is somewhat unexpected. It apparently stems from the 
fact that the structure of the convection velocity field produced by
	the 3D simulations is very different
	from simple isotropic Kolmogorov turbulence. The picture that
	emerges from these simulations consists of  large scale slow up-flows
	penetrated by relatively fast and very localized
	down-drafts that are coherent over a signifficant portion of the
	simulation box and persistent for extended periods of time.
	This is what causes the anisotropy and also seems to conspire
	to change the scaling with frequency, and make it relatively flat. 
	This makes our results appear closer to Zahn's prescription, which
	is coincidental, given the different physical assumptions. The
	question of what exactly is the reason for the shallower
	frequency dependence of the dissipation is of course a very
	interesting one. However, using a perturbative approach,
	limits us in our ability to answer it. To properly address
	this question one would need to create a consistent
	hydrodynamical simulation that allows for the perturbation
	velocity field to be put directly into the equations of motion
	and not treated by a perturbative approach after the fact.
	This would also address the question of whether the expansion
	is actually converging and if taking the first nonzero term is
	a good approximation, which is currently only our hope.

	This enhanced dissipation is in better % can you quantify this statement in any way 
	agreement with data on the
	circularization of the orbits of Sun-like main sequence stars, and the
	location of the instability strip as discussed earlier. We currently cannot  make
	any statements about the dissipation of p-modes, because those do not satisfy 
	the assumption of linearity and incompressibility of the perturbation
	velocity over the simulation box. However, we have used a
	solar 3D convection simulation which is consistent with the
	solar p-mode spectrum.\\

	Note that our approach here is more appropriate to tides raised by a planet on a
	slow (non-synchronized) star (Sasselov 2003). The problem of binary stars
	circularization will require a detailed treatment and understanding of
	the feedback on the convection zone. On the other hand, the tidal
	dissipation in fast-rotating fully-convective planets and stars might
	be dominated by inertial waves (Wu 2004ab, Ogilvie \& Lin 2004). They
	are sensitive to turbulent viscosity however, and the linear scaling
	has a strong effect on their dissipation (Wu 2004b). This issue deserves
	further study.

\end{document}